\newcommand{\D}{A}
\newcommand{\B}{B}
\newcommand{\Ea}{E_{\rm a}}

\newcommand{\hd}{h_{\rm d}}
\newcommand{\xid}{\xi_{\rm d}}

\newcommand{\kT}{k_{\rm B}T}

\newcommand{\cg}{c}
\newcommand{\tD}{\tau_{{\rm D}}}

\newcommand{\e}{\vec{s}}

\newcommand{\n}{\vec{n}}
\newcommand{\ei}{\vec{s}_{i}}
\newcommand{\ej}{\vec{s}_{j}}

\newcommand{\nii}{\vec{n}_{i}}
\newcommand{\nj}{\vec{n}_{j}}

\newcommand{\rij}{\vec{r}_{ij}}

\newcommand{\Gij}{{\bf G}_{ij}}
\newcommand{\I}{{\bf I}}

\newcommand{\la}{\left\langle}
\newcommand{\ra}{\right\rangle}
\newcommand{\raa}{\right\rangle_{\rm a}}

\newcommand{\case}[2]{{\textstyle \frac{#1}{#2}}}

\documentclass{epl}

\begin{document}

\title{
Relaxation time of weakly interacting superparamagnets
}

\author{
P. E. J{\"o}nsson and J. L. Garc\'{\i}a-Palacios
}

\institute{
Department of Materials Science, Uppsala University
\\
Box 534, SE-751 21 Uppsala, Sweden
}


\maketitle

\begin{abstract}
The relaxation time of weakly interacting classical spins is
calculated by introducing the averages of the local dipolar
field, obtained by thermodynamic perturbation theory, in a rigorous
expression for the single-spin thermoactivation rate in a weak but
arbitrarily oriented field.
At low temperatures the non-trivial dependence of the
superparamagnetic blocking on the damping coefficient, numerically
found by Berkov and Gorn, is reproduced by our model and interpreted
in terms of the deviations from uniaxial anisotropy associated to the
transversal component of the dipolar field acting on each spin.
\end{abstract}

\section{Introduction}

The study of single-domain magnetic particles has been an
active field of research since the pioneer work of Stoner and
Wohlfarth \cite{stowoh48}, who studied the hysteretic rotation of the
magnetisation over the magnetic-anisotropy energy barrier under
the influence of applied fields, and N{\'e}el \cite{nee49} who
predicted that at non-zero temperature the magnetisation can
surmount the energy barrier as a result of thermal agitation.
Important progress has been made since Brown \cite{bro63} derived the
Fokker--Plank equation for the probability distribution of spin
orientations, starting from the stochastic Landau--Lifshitz equation,
and calculated the relaxation time for uniaxial particles in a
longitudinal field.
Recent work on spins with  non-axially symmetric potentials
revealed \cite{garetal99} a large dependence of the relaxation time
on the damping coefficient $\lambda$ in the medium-to-weak damping
regime ($\lambda$ measures the relative importance of the precession
and the damping in the dynamics).
Experiments on individual nanoparticles \cite{weretal97TA}
analysed with accurate asymptotes of the relaxation time
\cite{cofetal98prl}, gave damping coefficients in that regime:
$\lambda \approx 0.05$--$0.5$.

Turning our attention from independent particles to systems of
interacting particles, the complexity of the problem increases
drastically, as it becomes a many-body problem with a long-ranged and
reduced symmetry interaction mechanism---the dipolar interaction.
The approach to study the relaxation time $\tau$ has so far been
based on how the energy barriers of the spins are modified by the
interactions \cite{dorbesfio88,mortro94}.
With the $\tau$ so obtained, one can study the effects of the
interaction on the superparamagnetic blocking (the maximum
in the dynamical response at the temperature where $\tau$ becomes of the
order of the observation time).
However, the barrier-based approach to determine $\tau$
corresponds to assuming $\lambda\to\infty$, so that dynamical
features such as the precession of the spins are disregarded, as
incisively noted by Berkov and Gorn \cite{bergor2001}.
Indeed, numerical integration of the stochastic Landau--Lifshitz
equation for weakly interacting systems revealed
\cite{bergor2001} non-trivial effects of finite damping on the
superparamagnetic blocking, such as enhanced shifts of the
temperature of the maximum of the dynamical response  and 
non-monotonic behaviour of its height
with the interaction strength.

In this article we apply thermodynamic perturbation theory (treating
the anisotropy energy exactly and the dipolar energy perturbatively)
to calculate the averages of the dipolar field produced at the
position of a given spin.
These averages are then introduced in a rigorous low-field expansion
of the single-spin relaxation rate, which can be combined with a
recently derived perturbative formula for the equilibrium
susceptibility
\cite{jongar2001} to provide a model for the dynamical susceptibility
of weakly interacting spins.
In the low temperature range, where the superparamagnetic blocking
takes place, our model recreates the non-trivial damping
dependence of the blocking found in Ref.\ \cite{bergor2001}.
The analytical treatment permits us to ascribe the observed features
to the deviations from uniaxial symmetry of the potential due to the
transversal components of the dipolar field, deviations which render
the relaxation time very sensitive to the damping.

\section{Low-field relaxation rate}

In weak fields, we can handle the field dependence of the relaxation
time $\tau$ by expanding the relaxation rate $\Gamma=1/\tau$ in
powers of the field components (for convenience we use the field in
temperature units
$\vec{\xi}=m\vec{\B}/\kT$, where $m$ is the magnetic moment).
As the spins  have inversion symmetry in the absence of the
field, the relaxation rate will not change if the field is reversed
(since $\Gamma$ accounts for jumps over the energy barrier in both
directions).
Therefore, $\Gamma$ should be an even function of $\vec{\xi}$, and
hence all odd order coefficients in the expansion must be identically
zero.
For spins with uniaxial anisotropy, the second order corrections can
be written in terms of the longitudinal $\xi_{\|}$ and transverse
$\xi_{\perp}$ components of the field with respect to the anisotropy
axis.
Using, for instance, the invariance of the relaxation rate upon
field reflection through the barrier plane, where $\xi_{\perp}$ does
not change but $\xi_{\|}$ is reversed, we see that no quadratic term
of the form $\xi_{\|}\xi_{\perp}$ can occur.
This leads to the following generic form for the relaxation
rate of uniaxial spins in the presence of a weak, arbitrarily
directed, field
%_____________________________
%_____________________________
\begin{equation}
\label{Gamma:expansion:uni:gen}
\Gamma
\simeq
\Gamma_{0}
\Big(
1
+
\cg_{\|}
\xi_{\|}^{2}
+
\cg_{\perp}
\xi_{\perp}^{2}
\Big)
\;,
\end{equation}
%_____________________________
%_____________________________
where $\Gamma_{0}$ is the zero-field relaxation rate and the
expansion is actually valid to third order.
The coefficients
$\cg_{\|}$ and $\cg_{\perp}$ will be determined 
by choosing special configurations in which they are
known (strictly longitudinal and transverse fields).

We shall restrict our attention to low temperatures where
the superparamagnetic blocking takes place.
Then, the zero-field relaxation rate is given by Brown's result
\cite{bro63}
%_____________________________
%_____________________________
\begin{equation}
\label{Gamma0}
\Gamma_{0}
=
\frac{1}{\tD}
\frac{2}{\sqrt{\pi}}
\sigma^{3/2}
e^{-\sigma}
\;,
\end{equation}
%_____________________________
%_____________________________
where $\tD$ ($\propto1/\lambda$) is the relaxation time of isotropic
spins, $\sigma=\D/\kT$ is the anisotropy barrier in temperature units,
and corrections of order $1/\sigma$ are disregarded due
to the low $T$ assumption.
Note that in this case (and also in a longitudinal field) the damping
parameter $\lambda$ only enters through $\tD$ and hence it only
matters to establish a global time scale.
In other words, the results for different $\lambda$ presented in units
of $\tD$ show complete dynamical scaling, and in this sense the
$\lambda$ dependence is said to be trivial.
Expanding the corresponding expression for $\Gamma$ in the presence
of a longitudinal field $\xi_{\|}$ \cite{bro63,aha69} (disregarding
again $1/\sigma$ terms), we find
%_____________________________
%_____________________________
\[
\Gamma(\xi_{\|},\xi_{\perp}=0)
\simeq
\Gamma_{0}
\Big(
1
+
\case{1}{2}
\xi_{\|}^{2}
\Big)
\;.
\]
%_____________________________
%_____________________________
Comparison with the general expansion (\ref{Gamma:expansion:uni:gen})
gives the longitudinal coefficient $\cg_{\|}=1/2$.

There is no general expression for the relaxation time in the
presence of a non-zero transverse field valid for all values of
the relevant parameters.
Nevertheless, Garanin {\em et al.} \cite{garetal99} have derived a low
temperature formula valid for weak transverse fields, which is
perfectly suited for our purpose of determining
$\cg_{\perp}$, namely
%_____________________________
%_____________________________
\[
\Gamma(\xi_{\|}=0,\xi_{\perp})
\simeq
\Gamma_{0}
\Big[
1
+
\case{1}{4}
F(\alpha)
\xi_{\perp}^{2}
\Big]
\;,
\qquad
F(\alpha)
=
1
+
2
(2\alpha^{2}e)^{1/(2\alpha^{2})}
\gamma
\left(
1+\frac{1}{2\alpha^{2}}
\,,\,
\frac{1}{2\alpha^{2}}
\right)
\;.
\]
%_____________________________
%_____________________________
Here $\alpha=\lambda\,\sigma^{1/2}$ and
$\gamma(a,z)
=
\int_{0}^{z} dt\, t^{a-1}\,e^{-t}$
is the incomplete gamma function.
Comparing with the expansion (\ref{Gamma:expansion:uni:gen}) one
gets the transverse coefficient $\cg_{\perp}=F/4$.

On gathering these results we finally get the desired expression for
the low temperature relaxation rate in weak fields
%_____________________________
%_____________________________
\begin{equation}
\label{Gamma:expansion:uni}
\Gamma
\simeq
\Gamma_{0}
\Big[
1
+
\case{1}{2}
\xi_{\|}^{2}
+
\case{1}{4}
F(\alpha)
\xi_{\perp}^{2}
\Big]
\;,
\end{equation}
%_____________________________
%_____________________________
which constitutes a straightforward generalisation of the formula of
Garanin {\em et al.} \cite{garetal99} to an arbitrary field
orientation.
The function $F$ takes into account without further approximations
the effects of the precession.
$F$ decreases towards $1$ for strong damping (where $\cg_{\|}$ and
$\cg_{\perp}$ are of the same order of magnitude), while $F$ grows as
$1/\lambda$ for weak damping, where the relaxation time turns to be
very sensitive to the damping (or to transverse fields).

Let us compare the rigorous Eq.\ (\ref{Gamma:expansion:uni}) with the
corresponding result of M{\o}rup and co-workers (see, for instance,
Eq.\ (6) in Ref.\ \cite{mortro94}), which for $1/\sigma\ll1$ can be
written as
%_____________________________
%_____________________________
\begin{equation}
\label{Gamma:expansion:morup}
\Gamma_{\mbox{\scriptsize M{\o}rup}}
\simeq
\Gamma_{0}
\Big(
1
-
\xi_{\|}
+
\case{1}{2}
\xi_{\|}^{2}
+
\case{1}{4}
\xi_{\perp}^{2}
\Big)
\;.
\end{equation}
%_____________________________
%_____________________________
This expression has the undesirable feature of a term linear in the
field, which can be attributed to accounting for the escape from
one of the potential wells only.
Actually, if we average over both wells the linear term cancels out
and the corrected formula equals  the overdamped
$\lambda\to\infty$ limit of Eq.\ (\ref{Gamma:expansion:uni})
(then $F\to1$).
This is natural, since in Ref.\ \cite{mortro94} the effects of the
field were only considered via barrier changes, so that no
gyromagnetic effects were included, and their result can only be
correct when the precession can be neglected ($\lambda\to\infty$).
Besides, as their linear term disappeared upon random anisotropy
averaging, their results will correspond to the limits overdamped
{\em plus} random anisotropy of those derived here from Eq.\
(\ref{Gamma:expansion:uni}).

\section{Averages of the dipolar field}

Let us consider a system of magnetoanisotropic spins coupled
via the dipole-dipole interaction.
The dipolar field at the position $\vec{r}_{i}$ of the spin $\ei$
created by all other spins is given by
%_____________________________
% DIPOLE-DIPOLE ENERGY
%_____________________________
\begin{equation}
\label{dipolar}
\vec{\B}_{i}
=
\frac{\mu_{0}m}{4\pi a^{3}}
\sum_{j}
\Gij\cdot \ej
\;,
\qquad
\Gij
=
\frac{1}{r_{ij}^{3}}
\left(3\,\hat{r}_{ij} \hat{r}_{ij}
-
\I
\right)
\;,
\end{equation}
%_____________________________
%_____________________________
where the term with $j=i$ is omitted, $m$ is the magnitude of the 
magnetic moment, and
in the dipolar tensor $\rij=\vec{r}_{i}-\vec{r}_{j}$ and
$\hat{r}_{ij}=\rij/r_{ij}$.
All lengths are measured in units of the characteristic length $a$,
which is defined in such a way that $a^3$ is the mean volume occupied
around each spin.
In a simple cubic arrangement $a$ is the lattice constant and
for nanoparticles of volume $V$ the volume concentration of
particles is $V/a^3$.
For notational simplicity we are assuming that the parameters
characterising the different spins are identical (it is immediate
to generalise the expressions for different anisotropy constants,
magnetic moments, volumes, etc.).

For spins with uniaxial anisotropy the total energy of the system
can be written as
%____________________________
% TOTAL ENERGY
%____________________________
\begin{equation}
-\beta E
=
\sigma
\sum_{i}(\ei\cdot\nii)^{2}
+
\xid
\sum_{i>j} \ei\cdot\Gij\cdot\ej
\;,
\end{equation}
%_____________________________
%_____________________________
where $\beta=1/\kT$, $\nii$ is the unit vector along the anisotropy
axis of the $i$th spin, $\sigma$ is the anisotropy barrier divided
by the thermal energy, and 
%_____________________________
% DIMENSIONLESS QUANTITIES
%_____________________________
\begin{equation}
\label{dimless}
\xid
=
\frac{\mu_{0}}{4\pi a^3}
\frac{m^{2}}{\kT}
\;,
\qquad
\hd
=
\frac{\xid}{2\sigma}
\;.
\end{equation}
%_____________________________
%_____________________________
The quantity $\hd$ is a convenient temperature independent measure
of the interaction strength ($\propto$ concentration) equal to the
magnitude of the field, measured in units of the maximum anisotropy
field $\B_{K}=2\D/m$, produced at a given position by a spin located
at a distance $a$.

Using thermodynamic perturbation theory (Ref.\
\cite{lanlif5}, \S~32) to expand the Boltzmann distribution
$W=Z^{-1}\exp (-\beta E)$ in powers of $\xid$, one gets an
expression of the form
%_____________________________
% GIBBS DISTRIBUTION: APPROXIMATE
%_____________________________
\begin{equation}
\label{Boltzmann}
W(\ei)
=
W_{\rm a}(\ei)
\left[
1
+
\xid
w_{1}(\ei)
+
\cdots
\right]
\;.
\label{Wexp}
\end{equation}
%_____________________________
%_____________________________
Here $W_{\rm a}=\prod_{i}Z_{\rm a}^{-1}\exp(-\beta\Ea)$ is the
Boltzmann distribution of the non interacting ensemble (in our case
$\Ea$ includes the magnetic anisotropy) and $w_{1}(\ei)$ is
linear in the dipolar energy (and hence quadratic in the spins).
The calculation of the observables reduces to computing averages
weighted by the non interacting probability distribution
$\la\cdots\raa
=
Z_{\rm a}^{-1}
\int\!
\frac{d^{2}\e}{2\pi}
\exp(-\beta\Ea)
(\cdots)$
of low grade powers of the spin variables.
In this way, the results are exact in the magnetic anisotropy
and only perturbational in the dipolar interaction.

The averages of terms linear and quadratic in $\e$, weighted by the
noninteracting distribution, can be calculated by means of the
following ``algorithms" \cite{jongar2001}
%_____________________________
% algorithms
%_____________________________
\begin{equation}
\label{alg}
\la
(\e\cdot\vec{v}_{1})
\raa
=
0,
\qquad\la (\e\cdot\vec{v}_{1})(\e\cdot\vec{v}_{2})\raa
=
\case{1}{3}
\left(1-S_{2}\right) \, \vec{v}_{1}
\cdot\vec{v}_{2}
+
S_{2} \, (\n\cdot\vec{v}_{1})(\n\cdot\vec{v}_{2})
\;,
\end{equation}
%_____________________________
%_____________________________
where $\vec{v}_{1}$ and $\vec{v}_{2}$ are arbitrary constant vectors.
The first average vanishes (actually any odd power) since the magnetic
anisotropy has inversion symmetry [$W_{\rm a}(-\ei)=W_{\rm a}(\ei)$]
and there is no external bias field.
The quantity $S_{2}$ is the average of the second order Legendre
polynomial $S_{2}=\langle\frac{1}{2}[3(\e\cdot\n)^{2}-1]\rangle_{\rm
a}$, and it can be written in terms of the one-spin partition function
$Z_{\rm a}$, as
%_____________________________
% S_{2} & PARTITION FUNCTION
%_____________________________
\begin{equation}
\label{S2&Z}
S_{2}(\sigma)
=
\frac{3}{2}
\left(
\frac{e^{\sigma}}{\sigma Z_{\rm a}}
-
\frac{1}{2\sigma}
\right)
-
\frac{1}{2}
\;,
\qquad
Z_{\rm a}
=
\sqrt{\pi/\sigma}
\,{\rm erf}({\rm i}
\sqrt{\sigma})
\;.
\end{equation}
%_____________________________
%_____________________________

In the equations for the relaxation rate the field enters squared
[Eq.~(\ref{Gamma:expansion:uni})].
Therefore, we do not calculate the statistical mechanical average of
$\vec{\B}_{i}$ [Eq.\ (\ref{dipolar})] and plug it into $\Gamma$
(which would be a sort of mean-field approach) but we average instead
the combinations of field variables as they enter in the expression
for $\Gamma$.
Since $\B^{2}_{\perp}=\B^{2}-\B^{2}_{\|}$, we average, by means of the
algorithms (\ref{alg}), the square of the field and the square of the
projection along the local anisotropy axis
$\B_{i,\|}^{2}
=
(\vec{\B}_{i}\cdot\nii)^{2}$, getting ($\xi_{i}=mB_{i}/\kT$)
%_____________________________
% H2 para
% H2 perp
%_____________________________
\begin{eqnarray}
\label{h:para}
\big\langle
\xi_{i,\|}^{2}
\big\rangle
&=&
\frac{\xid^{2}}{3}
\sum_{j}
\big[
\left(1-S_{2}\right)
\,
(\nii\cdot\Gij\cdot\Gij\cdot\nii)
+3S_{2} \, (\nii\cdot\Gij\cdot\nj)^{2}\big]
\;,
\\
\label{h:perp}
\big\langle
\xi_{i,\perp}^{2}
\big\rangle
&=&
\frac{\xid^{2}}{3}
\sum_{j}
\big[
6r_{ij}^{-6}
+
3S_{2} \, r_{ij}^{-3} (\nj\cdot\Gij\cdot\nj)
\nonumber\\
& &
\qquad
\quad
{}
-\left(1-S_{2}\right)
\,
(\nii\cdot\Gij\cdot\Gij\cdot\nii)
-3S_{2} \, (\nii\cdot\Gij\cdot\nj)^{2}\big]
\;.
\end{eqnarray}
%_____________________________
%_____________________________
The averages have been calculated to order $\xid^{2}$ (so we just need
Eq.\ (\ref{Boltzmann}) to zero order; $w_{1}$ only enters in the
third order corrections).
Note that the averaged fields depend on $T$ via $S_{2}(\sigma)$,
reflecting the fact that the field created at a given position can be
different if, for example, the source spins are almost freely
rotating (high $T$, $S_{2}\to0$) or are almost parallel to their
anisotropy axes (low $T$, $S_{2}\to1$).

One may wonder about the validity of these field averages below the
superparamagnetic blocking, where the spins are not in complete
equilibrium.
However, since at those temperatures the spins are still in
quasi-equilibrium confined to one of the two wells, we can repeat the
derivation of the algorithms (\ref{alg}) restricting the phase space
for integration to one well.
In this case, averages of the form $\la\e\cdot\vec{v}_{1}\raa$ do not
vanish, and should be considered together with
$\la (\e\cdot\vec{v}_{1})(\e\cdot\vec{v}_{2}) \raa$, which being even
in $\e$ is not modified.
The extra terms associated with $\la \e\cdot\vec{v}_{1} \raa$,
however, vanish if the overall state is demagnetised, and we recover
Eqs.\ (\ref{h:para})--(\ref{h:perp}).
This can be interpreted as if two nearby blocked spins, one
in its upper well and the other in the lower, create a net field at
the position of a third spin similar to that created by any of them
when in equilibrium, since each of the spins approximately compensates
for the restricted phase space of the other.

The general expressions for the longitudinal and transversal fields
are notably simplified in some important situations.
For a system with parallel anisotropy axes (e.g., in a single crystal
of magnetic molecular clusters, or ferrofluids frozen in a strong
magnetic field) we equate all the $\nj$ to $\n$.
For a system with randomly distributed anisotropy axes we replace
expressions involving $f(\nj)$ by integrals
$\int d^{2}\n\,f(\n)\equiv\overline{f}$, and use
$\overline{(\n\cdot\vec{v}_{1})(\n\cdot\vec{v}_{2})}
=
\frac{1}{3}\vec{v}_{1}\cdot\vec{v}_{2}$.
In both cases the final expressions involve some rapidly convergent
sums over the lattice.
Let us concentrate in cases of ``sufficiently isotropic" lattices, in
the sense of fulfilling $\sum (r_x)^k=\sum(r_y)^k=\sum(r_z)^k$, e.g.,
cubic and completely random lattices (incidentally, the type of
arrangements for which in the classical Lorentz cavity field calculation the
contribution of the dipoles inside the ``small" sphere vanishes).
	Similarly we consider large enough systems, so that all spins
have approximately equivalent surroundings (then the index $i$ on the
different quantities can be dropped).
	Under these circumstances the ``lattice sums" involved are
%_____________________________
%_____________________________
\begin{eqnarray}
\label{R&T}
{\cal R}
=
2
\sum_{j}
r_{ij}^{-6}
\;,
\qquad
{\cal T}
=
\sum_{j}
(\n\cdot\Gij\cdot\n)^{2}
\;,
\end{eqnarray}
%_____________________________
%_____________________________
where the terms with $j=i$ are of course omitted.

For a system with aligned anisotropy axes the averaged fields
are given in terms of the lattice sums (\ref{R&T}) by the compact
expressions
%_____________________________
% H2 para para + symmetic lattice
% H2 perp para + symmetic lattice
%_____________________________
\begin{equation}
\label{h:para:perp:aligned}
\big\langle
\xi_{\|}^{2}
\big\rangle
=
\frac{\xid^{2}}{3}
\left[
\left(1-S_{2}\right)
{\cal R}
+
3S_{2}\,{\cal T}
\right],
\qquad
\big\langle
\xi_{\perp}^{2}
\big\rangle
=
\frac{\xid^{2}}{3}
\left[
(2+S_{2}) {\cal R}
-
3S_{2}\,{\cal T}
\right]
\;,
\end{equation}
%_____________________________
%_____________________________
while for randomly distributed anisotropy axes they read
%_____________________________
%_____________________________
\begin{equation}
\label{h:para:perp:random}
\overline{
\big\langle
\xi_{\|}^{2}
\big\rangle
}
=
\frac{\xid^{2}}{3}
{\cal R},
\qquad
\overline{
\big\langle
\xi_{\perp}^{2}
\big\rangle
}
=
\frac{\xid^{2}}{3}
2{\cal R}
\;.
\end{equation}
%_____________________________
%_____________________________
In both cases
$\la\xi^{2}\ra
=
\big\langle\xi_{\|}^{2}\big\rangle
+
\la\xi_{\perp}^{2}\ra$
is simply given by the temperature independent result
$\la\xi^{2}\ra=\xid^{2}{\cal R}$.
This independence also holds for the field components after
the random anisotropy average [Eq.\ (\ref{h:para:perp:random})], but
not for parallel axes [Eq.\ (\ref{h:para:perp:aligned})] where the
field components remain $T$ dependent.

\section{Superparamagnetic blocking}

The dependence of the features of the superparamagnetic blocking on
the interaction strength $\hd=\xid/2\sigma$ has been a subject of some
controversy.
Two main approaches \cite{dorbesfio88,mortro94} addressed this problem
on the basis of the modifications of the energy barriers by the
interactions.
However, this type of approach overlooks the fact that not only the
energy landscape is important, but also how the spin evolves in it,
depending on if the spin is precessing almost freely (weak damping) or
strongly damped.
Indeed, Berkov and Gorn \cite{bergor2001} have shown with rigorous
Langevin dynamics simulations that for weak interactions the
position of the blocking temperature
$T_{\rm b}$ (where
$\chi''$ reaches its maximum) of strongly damped spins is hardly
affected by the interaction strength, whereas for weak damping
$T_{\rm b}$ significantly decreases with
$\hd$ and the peak height behaves non-monotonically.
The apparent discrepancy with some experimental results, in which
$T_{\rm b}$ increases with $\hd$, was attributed in Ref.\
\cite{bergor2001} to the different behaviour of systems with weak
anisotropy (or moderate-to-high interactions), where the energy barriers are
mostly due to the interactions and hence grow with $\hd$.

In order to assess the features of the superparamagnetic blocking
emerging from our model, we make use of a rigorous perturbative
expansion of the equilibrium susceptibility $\chi_{\rm eq}$ in powers
of $\xid$ (to second order, see Ref.\ \cite{jongar2001}).
This expression, together with the relaxation rate $\Gamma$ obtained
when the averaged fields
(\ref{h:para:perp:aligned})--(\ref{h:para:perp:random}) are
introduced in Eq.\ (\ref{Gamma:expansion:uni}), can be combined in a
simple Debye-type formula
%_____________________________
%_____________________________
\[
\chi
=
\chi_{\rm eq}
\,
\frac{\Gamma}{\Gamma+{\rm i}\,\omega}
\;.
\]
%_____________________________
%_____________________________
Naturally, at low enough temperatures \cite{jongar2001} the results
will become invalid by the very nature of the approximations
involved, although this does not affect the characteristics of
the superparamagnetic blocking for weak interactions.

\begin{figure}
\onefigure[width=14cm]{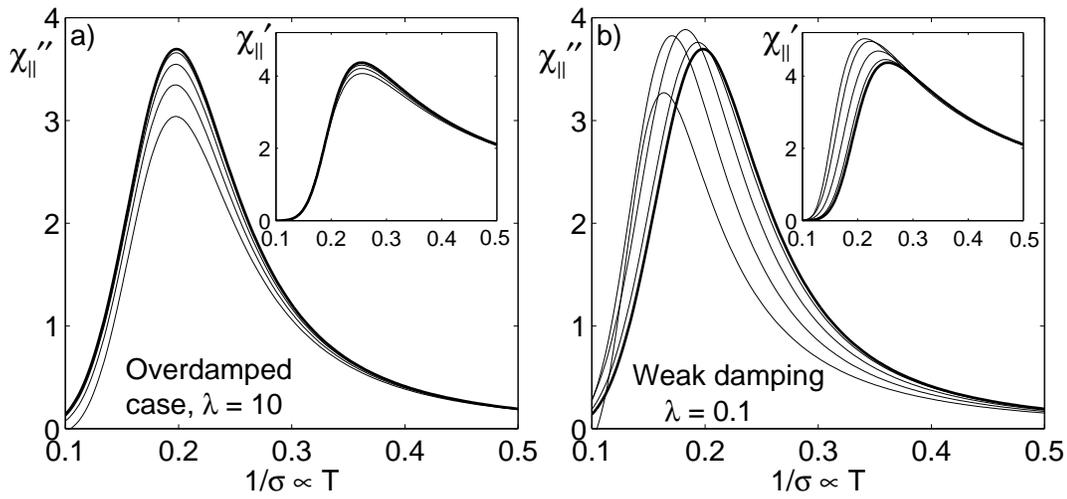}
\caption[]{
Imaginary component of the dynamical susceptibility vs.\
temperature (the real component is shown in the inset) 
for a spherical sample and spins placed in a simple cubic lattice.
The anisotropy axes are all parallel and the response probed along
their common direction.
The dipolar interaction strength $\hd=\xid/2\sigma$ is: $\hd=0$
(thick lines), 0.004, 0.008, 0.012, and 0.016 from (a) up to down and
(b) right to left.
The frequency is $\omega\tD/\sigma=2\pi\times0.003$.
}
\label{Fig: fig1}
\end{figure}

The dynamical response for a large spherical sample with parallel
anisotropy axes and simple cubic lattice structure is shown in
Fig.~\ref{Fig: fig1} (the lattice sums required are ${\cal R}=16.8$
and ${\cal T}=13.4$).
In the overdamped case, $T_{\rm b}$ is not noticeably affected by the
dipolar interaction and the height of the susceptibility
peak decreases monotonically with $\hd$.
This corresponds to the effect of a slight decrease of $T_{\rm b}$
found in Ref. \cite{mortro94} for random anisotropy, which is
observable only at very high frequencies (e.g., M\"ossbauer).
For weak damping, however, Fig.~\ref{Fig: fig1} shows that the
blocking temperature significantly decreases as the interaction
strength increases and, in addition, the peak height of $\chi''$
initially rises for small values of $\hd$ and then decreases for
larger values.
The same behaviour can be seen in $\chi'$ with the appropriate choice of parameters.

The features shown in Fig.~\ref{Fig: fig1} are in complete
agreement with the results obtained by Berkov and Gorn
\cite{bergor2001}.
The analytical treatment employed here, however, allows us to readily
trace back the origin of the results obtained and to interpret them in
terms of the different $\hd$ dependences of $\Gamma$ and $\chi_{\rm
eq}$.
For overdamped systems the coefficients $\cg_{\|}=1/2$ and
$\cg_{\perp}=F/4\simeq1/4$ in the expression
(\ref{Gamma:expansion:uni}) for the relaxation rate are of order
unity and lead to a slight increase of $\Gamma$ (decrease of $T_{\rm
b}$) with $\hd$, while the entire curve is lowered by the reduction of
$\chi_{\rm eq}$ with $\hd$.
For weak damping, however, $F\propto1/\lambda$ is large and makes
$\Gamma$ very sensitive to the interaction strength, moving the peak
quickly towards low $T$ as soon as $\hd$ departs from zero.
Then, as the coefficients in $\chi_{\rm eq}$ are not as large as
$\cg_{\perp}$, the initial decrease of the equilibrium susceptibility
is smaller than the increase associated to the quick shift of the
blocking to low $T$, where $\chi_{\rm eq}$ is higher (roughly
$\propto1/T$), producing the rise of the peak.
If $\hd$ is further increased the decrease of $\chi_{\rm eq}$ starts
to be competitive and the peaks are reduced as they shift to
lower temperatures.

The different behaviours are related to the presence of transverse
components of the local fields, which create a saddle point in the
uniaxial potential barrier of the spins, turning the relaxation rate
sensitive to $\lambda$ \cite{garetal99}.
We can picture the underlying physical mechanism as follows
\cite{garsve2000}.
Consider one spin that after a ``favourable" sequence of fluctuations,
reaches a point close to the top of the barrier but does not surmount
it.
In the subsequent spiralling down back to the bottom of the potential
well, a strongly damped spin descends almost immediately, whereas a
weakly damped spin executes several rotations ($\sim1/\lambda$) about
the anisotropy axis.
This allows the latter spin to pass close to the saddle area, where
it will have additional opportunities, not available for the damped
spin, to cross the barrier.
As we see, this mechanism duly combines the characteristics of the
potential {\em and\/} the dynamical evolution of the
spins in the potential.
Note finally that the transverse components are non-zero even for
parallel anisotropy axes [Eq.\ (\ref{h:para:perp:aligned})], so there
is no need, as in the case of non-interacting particles, to  
appeal to oblique fields, applied \cite{garetal99} or
probing \cite{garsve2000}, 
to find a large sensitivity to the damping in interacting systems.

\section{Summary and Conclusions}

We have proposed a model for the relaxation time of weakly
interacting superparamagnets.
The single-spin relaxation time at low fields is generalised for
an arbitrary directed field, and the components of the local field
calculated by thermodynamical perturbation theory.
The non-monotonic behaviour of the height of the dynamical
susceptibility peak with the interaction strength and the increased 
lowering of the blocking temperature with decreasing damping, 
discovered numerically by
Berkov and Gorn, are captured by our model.
These features are interpreted in terms of the different sensitivity,
depending on the damping strength, of the relaxation rate to
transverse fields, which for interacting spins are provided by the
dipolar interaction.

\acknowledgments
We thank D. V. Berkov and N. L. Gorn for sending us a preprint of
their article.
This work was financially supported by The Swedish Natural Science
Research Council (NFR).

%\bibliography{PJref}

\end{document}